\documentclass[journal]{IEEEtran}
\pdfoutput=1 
\usepackage{ifpdf}
\usepackage{cite}
  \ifCLASSINFOpdf
     \usepackage[pdftex]{graphicx}
  \else
    \usepackage[dvips]{graphicx}
  \fi
\usepackage{amsmath}
\usepackage{algorithmic}
\usepackage{array}
   \ifCLASSOPTIONcompsoc
    \usepackage[caption=false,font=normalsize,labelfont=sf,textfont=sf]{subfig}
   \else
   \usepackage[caption=false,font=footnotesize]{subfig}
   \fi
\usepackage{url}
\usepackage{bm}
\begin{document}
\title{Modeling and characterization of a rectangular waveguide grating structure using transmission line theory for planar Cerenkov masers}
\author{Ye~Chen
and~Yaogen~Ding
% <-this % stops a space
\thanks{Y. Chen was with the University of Chinese Academy of Sciences and now is with Deutsches Elektronen-Synchrotron DESY. E-mail address: ye.lining.chen@desy.de.}% <-this % stops a space
\thanks{Y.-G. Ding is with the Chinese Academy of Sciences.}% <-this % stops a space
}
\markboth{Article}%
{Chen \MakeLowercase{\textit{et al.}}: Modeling and characterization of a rectangular waveguide grating structure using transmission line theory for planar Cerenkov masers}
\maketitle

%%*************************************************************************
\begin{abstract}
A modeling approach is proposed based on transmission line theory for the characterization of the periodic rectangular waveguide grating (RWG) structure. Using an equivalent circuit (EC) model the dispersion equation of the structure is derived with largely reduced workloads as compared to the conventional field-theory method. An EC based analysis of the RWG structure is performed. Numerical results show a good consistency between the two methods as varying structural parameters of significance. The proposed approach is also used for the taper design with the objective of minimizing wave reflection of the structure. A resulting multistage taper can deliver a low cumulative reflection coefficient on the order of $10^{-3}$. Furthermore, the coherence performance of an RWG based planar Cerenkov maser (PCM) is studied on the driving electron beam interacting with the traveling harmonic wave. This includes the impacts of the grating height uniformity, due to practical machining uncertainty, on the net wave reflection as well as on the growth rate of the wave in the maser. The obtained results show, that a non-uniformity on the order of 50 $\mu m$ in the grating height can increase the reflection level by at least one order of magnitude. The PCM coherence can be considerably degraded, in terms of a significant reduction in the wave growth rate of more than 30$\%$ with respect to its theoretical value.
\end{abstract}

%%*************************************************************************

\IEEEpeerreviewmaketitle

%%*************************************************************************
\section{Introduction}
\IEEEPARstart{P}{lanar} Cerenkov maser (PCM), as a promising electromagnetic radiation source, has drawn growing interests in the millimeter (mm) and / or sub-mm wavelength range over the last decades\cite{Collin, Dave, Walsh, Joe, Chang, Lee, Mehrany, McVey, Garate}. A typical radiation scheme of the PCM takes a rectangular waveguide periodically lined with metallic gratings to slow down the phase velocity of the traveling wave in the maser, such that a passing-by relativistic planar electron beam can be synchronized with the wave under the synchronous condition of velocity matching. The kinetic energy of the electron beam can thus be converted to the electromagnetic field energy by means of the so-called beam-wave interaction process. 

Adoption of a rectangular waveguide grating (RWG) structure in the PCM takes advantages of the facts, in comparison to dielectric-lined slow-wave structures, that the metallic grating shows better tolerance to electron interception and stronger capability in heat dissipation. This is of particular importance for high power applications. However, the PCM performance is sensitive to geometrical uniformity of periodic gratings. This refers, especially, to a slight variation of the grating height along the maser with respect to its designed value, due to practical machining error within certain accuracy. A very important role plays this effect on the coherence performance of the PCM for practical applications.

Design of an RWG based PCM usually starts with characterization of the RWG structure. This process refers to deriving the dispersion equation of the structure as well as finding solutions to the equation. This can be done by modeling a field matching problem with continuity conditions applied on the interface between the grating groove inside and the region above the grating\cite{Collin, Joe, McVey, Chen2011}. Solving the equation gives relations between the working frequency and the propagation constant of the guided wave for electromagnetic modes on different orders. This field-theory method principally is rigorous, but requires rather lengthy derivations. And yet, mathematically building up a physical model, in terms of dividing the whole calculation domain into multiple field regions with proper boundary conditions, may not be trivial for RWG-like structures with even more geometrical and / or dielectric features\cite{Cao20151, Cao20152}. 

In this paper, the modeling of the RWG structure is done using transmission line theory instead of the conventional field theory. A simple equivalent circuit (EC) model is proposed in Section \ref{modeling} to derive the dispersion equation. In Section \ref{validation}, numerical simulations of the dispersion relation are carried out to validate the proposed method by varying crucial geometrical parameters of the structure. A physical interpretation of the approach is given on the basis of the consistency with the field-theory method. Moreover, the proposed approach is extended in Section \ref{taper} for the taper design of the PCM. In Section \ref{impact}, the coherence performance of the PCM is investigated taking into account geometrical non-uniformities of the gratings. A summary and an outlook are given in Section \ref{conclusion}.

%%*************************************************************************
\section{Equivalent Circuit Based Modeling}\label{modeling}
\begin{figure}[!t]
\centering
\includegraphics[width=80mm]{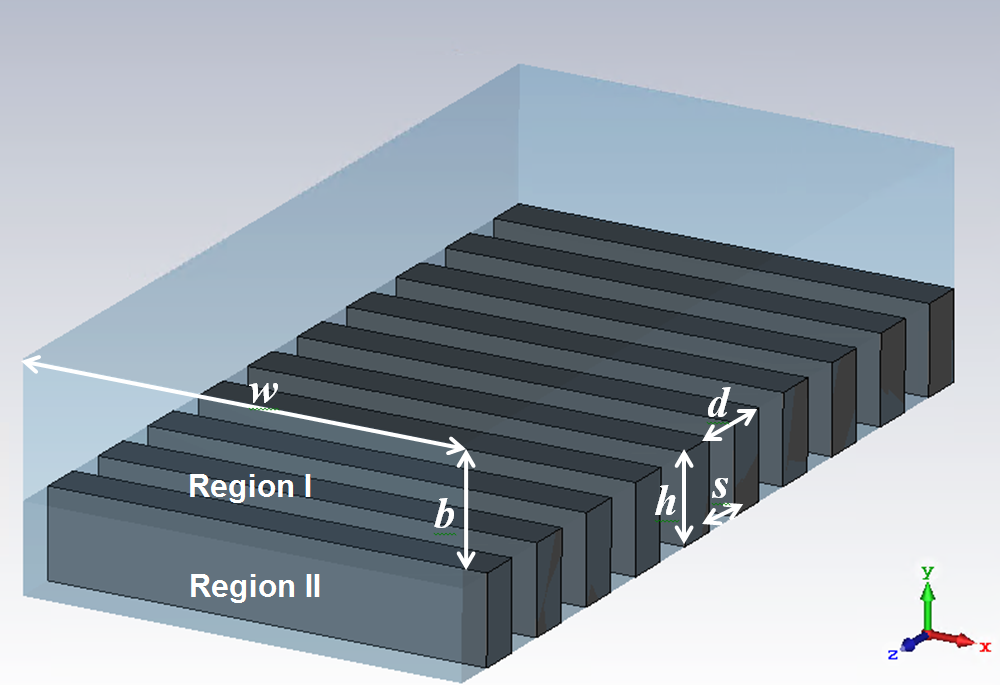}
\caption{Sketch of a rectangular waveguide grating (RWG) structure.}
\label{fig:1}
\end{figure}

Figure \ref{fig:1} shows a sketch of the RWG structure. Introducing periodic gratings into the smooth rectangular waveguide enables the propagation of the wave traveling on the surface of the grating with a phase velocity lower than the speed of light. The local fields within the grating groove are usually treated, in a conventional way, as TEM standing waves when the groove width $s$ is smaller than the wavelength. The role of the grating can be considered as storing the local electromagnetic energy. Each grating unit is then equivalent to a capacitive or inductive circuit component with energy storage. The periodic grating structure can thus be interpreted as the cascade of the equivalent microwave networks of all individual circuit components. The geometrical and electrical properties of the RWG structure can be easily analyzed using the simple circuit model, instead of using intricate field formalisms.
  
\begin{figure}[!t]
\centering
\includegraphics[width=80mm]{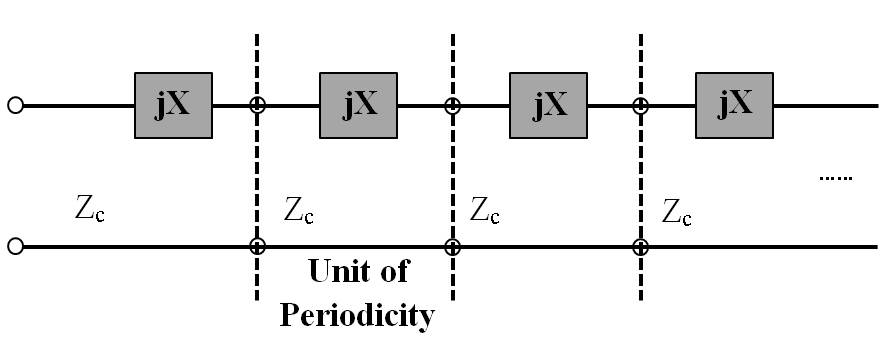}
\caption{Representation of the RWG periodic structure by cascaded equivalent circuit units.}
\label{fig:2}
\end{figure}

In Fig. \ref{fig:1}, the geometrical parameters are defined with $h$, $s$, $d$, $b$ and $w$ as the height, width, period of the grating, a distance of the grating surface to upper boundary of the waveguide and the transverse width of the waveguide, respectively. As shown, each period of the structure consists of Region I (above grating) and Region II (inside grating). For an equivalent circuit (EC) analysis, the Region I can be represented as parallel plate transmission line with a characteristic impedance of $Z_c$ = $Z_0$$\textit{b}$ per unit width, where $Z_0$ denotes the characteristic impedance of the line. The Region II is then considered as a shorted branch of the line periodically connected to the main line with a period of $\textit{d}$. This shorted branch exhibits a characteristic reactance of $\textit{jX}$ (= $jZ_0stank_0h$\cite{Collin}) to the main line. Thus, each grating period can be modeled as an EC unit consisting of three cascaded two-port microwave networks. This includes, specifically, a segment of transmission line with an electrical length of $\textit{l}$, a serial reactance of $\textit{jX}$, and another segment of line in the same length. The whole periodic structure can be considered as the cascade of all the EC units (see Fig. \ref{fig:2}). 

Using the proposed EC model, the wave propagation in the RWG structure can be described using a transfer matrix $\bm{A_t}$. Given the normalized characteristic impedance of the serial reactance $\overline{X}$ = $X$/$\textit{$Z_c$}$, the electrical length $\textit{l}$ as $k_0$$\textit{d}/2$ with $k_0$ denoting the wave number in free space, the overall transfer matrix of each EC unit is expressed as
\begin{equation} 
\boldsymbol{A_t}
=
\prod\limits_{i=1}^3\boldsymbol{A_i}
=
\begin{bmatrix}
cos l & jsin l \\
jsin l & cos l
\end{bmatrix}
\begin{bmatrix}
1 & $\textit{j$\overline{X}$}$ \\
0 & 1
\end{bmatrix}
\begin{bmatrix}
cos l & jsin l\\
jsin l & cos l
\end{bmatrix},\label{eq:1}
\end{equation}
where $\boldsymbol{A_i}$ represents the transfer matrix of the three cascaded EC elements, respectively. This directly gives
\begin{equation} 
\boldsymbol{A_t}
=
\begin{bmatrix}
cos2l-\frac{\overline{X}}{2}sin2l& j(sin2l+\frac{\overline{X}}{2}cos2l+\frac{\overline{X}}{2})\\
j(sin2l+\frac{\overline{X}}{2}cos2l-\frac{\overline{X}}{2})&cos2l-\frac{\overline{X}}{2}sin2l
\end{bmatrix}.\label{eq:2}
\end{equation}
Applying conditions of periodicity leads to
\begin{equation}
\begin{bmatrix}
\textit{$\boldsymbol{V_n}$}\\
\textit{$\boldsymbol{I_n}$}
\end{bmatrix} 
=
\textit{$\boldsymbol{A_t}$}
\begin{bmatrix}
\textit{$\boldsymbol{V_{n+1}}$}\\
\textit{$\boldsymbol{I_{n+1}}$}
\end{bmatrix}\label{eq:3}
\end{equation}
and
\begin{equation}
\begin{bmatrix}
\textit{$\boldsymbol{V_n}$}\\
\textit{$\boldsymbol{I_n}$}
\end{bmatrix} 
=
\begin{bmatrix}
e^{\gamma d}  & 0\\
0 & e^{\gamma d}
\end{bmatrix}
\begin{bmatrix}
\textit{$\boldsymbol{V_{n+1}}$}\\
\textit{$\boldsymbol{I_{n+1}}$}
\end{bmatrix}. \label{eq:4}
\end{equation}
where $\gamma$ represents wave propagation constant and $\boldsymbol{V(/I)_{i}}$ denotes the representative voltage (/current) wave at the $i_{th}$ port of the periodic structure.

Combining Eqs. (\ref{eq:2})-(\ref{eq:4}) defines an eigenvalue equation of $\gamma$ as
\begin{equation}
\textit{$\boldsymbol{D}$}\\
\begin{bmatrix}
\textit{$\boldsymbol{V_{n+1}}$}\\
\textit{$\boldsymbol{I_{n+1}}$}
\end{bmatrix}
=0 \label{eq:5}
\end{equation}
with
\begin{equation}
\textit{$\boldsymbol{D}$}\\
=
\begin{bmatrix}
cos2l-\frac{\overline{X}}{2}sin2l-e^{\gamma d} &  j(sin2l+\frac{\overline{X}}{2}cos2l+\frac{\overline{X}}{2})\\
j(sin2l+\frac{\overline{X}}{2}cos2l-\frac{\overline{X}}{2})  & cos2l-\frac{\overline{X}}{2}sin2l-e^{\gamma d}
\end{bmatrix}.
\label{eq:6}
\end{equation}

Only if the determinant value of Eq. (\ref{eq:6}) equals zero, solutions of $\boldsymbol{V_{n+1}}$ and $\boldsymbol{{I_{n+1}}}$ exist. This leads to the dispersion equation of the RWG structure in cold-state as 
\begin{equation}
cosh\gamma d=cos2l-\frac{\overline{X}}{2}sin2l
\label{eq:7}
\end{equation}
with the characteristic reactance of the shorted branch $j$$\overline{X}$ = $j$$\frac{s}{b}$$tank_0 h$, which exhibits inductive and capacitive to the main transmission line for the working frequencies satisfying 0 $<$ $k_0$$h$ $<$ $\pi$/2 and $\pi$/2 $<$ $k_0$$h$ $<$ $\pi$, respectively. As $k_0$$h$ $=$ $\pi$/2, the shorted branch resonates and approximately defines the first critical frequency. An increase of the grating height from the this point on would lead to a further reduction of the phase velocity in the first passband. Detailed characteristics are shown in the following section. 

%%*************************************************************************
\section{Validation of Method}\label{validation}
\begin{figure}[!htb]
\centering
\includegraphics[width=85mm]{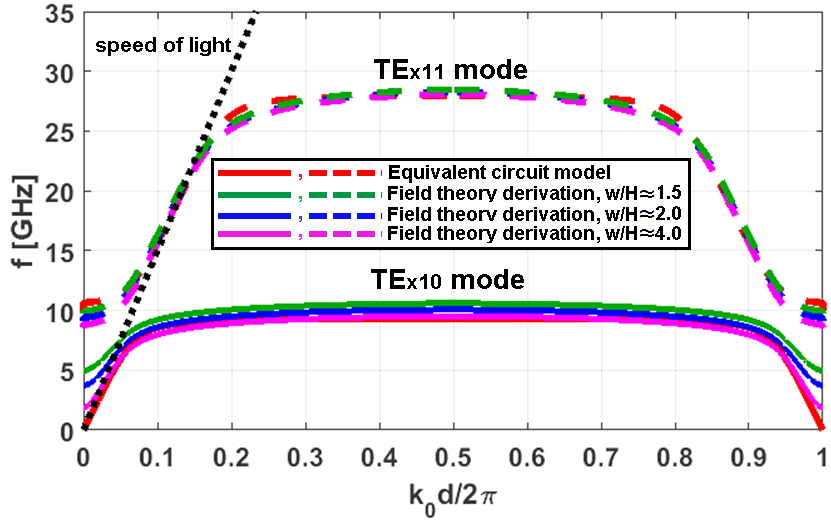}
\caption{Comparison of dispersion relations between the EC method and field-theory method by varying aspect ratios of the waveguide. $H$ (= $b$ + $h$): waveguide height. Following the mode naming convention, the mode $TE_{xlm}$ refers to a TE mode with respect to the $x$ direction with $l$ as the mode number and $m$ as the sequential number of the modes on different orders.}
\label{fig:3}
\end{figure}

The dispersion relations obtained from Eq. (\ref{eq:7}) are illustrated in Fig. \ref{fig:3} for the first two lowest-order modes, $TE_{x10}$ and $TE_{x11}$ (transverse electric mode w.r.t. x). The results are compared to those computed from the field-theory method following the derivations in Refs. \cite{Joe, McVey, Zhao2011} and \cite{Chen2011}. To demonstrate the influence of the waveguide edge effect on the dispersion relation, the comparisons are conducted by means of varying the aspect ratio of the waveguide from 1.5 to 4.0. The structural parameters used in the calculation are displayed in Table \ref{tab:1}. The dashed line in Fig. \ref{fig:3} represents the light line. As shown, in the slow-wave regime, the dispersion tendencies described by the field theory (green, blue and pink curves) are in agreements with those obtained from the equivalent circuit (EC) method (red curves) for both modes. As the aspect ratio of the waveguide is larger than 2, the dispersion relations derived by the EC method excellently fit the field theory. Under this condition the edge effect of the waveguide does not significantly contribute to the results. 

\begin{table}[!t]
\begin{center}
\caption{Geometrical parameters of the RWG structure.}\label{tab:1}
\begin{tabular}{llr}
\hline
Parameter    & Value & Unit\\
\hline
height above grating, \textit{b}      & 10      &mm   \\
height of grating groove, \textit{h}  & 8       &mm   \\
waveguide width, \textit{w}           & 20-80   &mm   \\
width of grating groove, \textit{s}      & 1       &mm   \\
grating period, \textit{d}            & 1.5-8   &mm   \\
\hline
\end{tabular}
\end{center}
\end{table}

\begin{figure}[!htb]
\centering
\includegraphics[width=85mm]{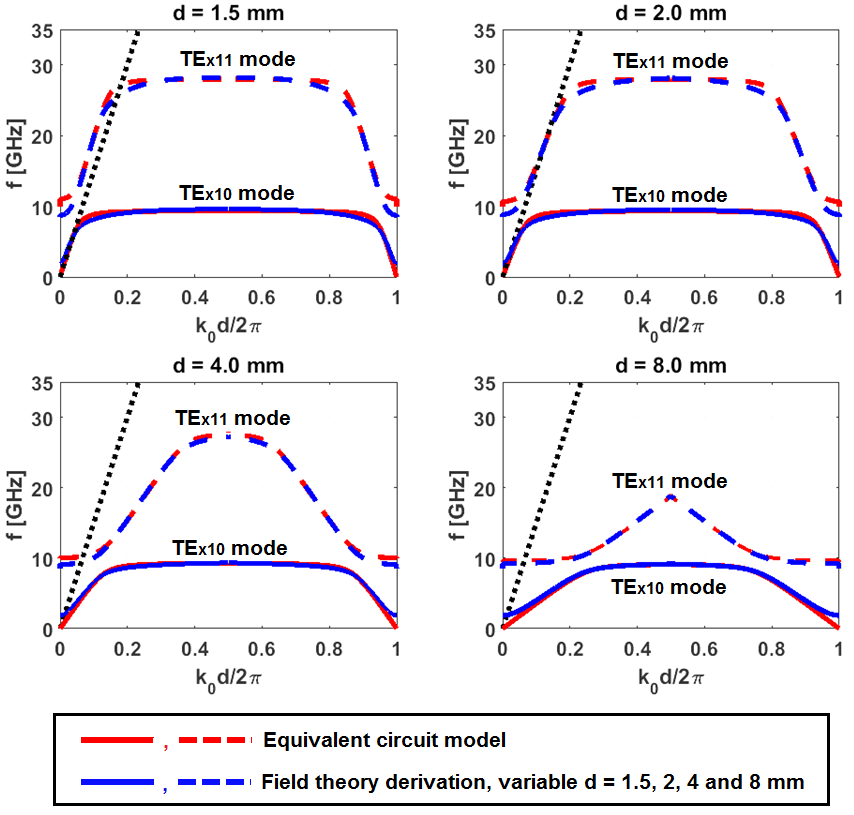}
\caption{Dispersion relations versus grating period $d$ as the aspect ratio of the waveguide $w$/$H$ = 4.}
\label{fig:4}
\end{figure}

Figure \ref{fig:4} shows another characteristic calculation of the dispersion relation using various grating periods. The results are, once again, compared to the field-theory method. It can be seen, that good consistencies are still retained between the two methods for a large range of the grating period $d$ varying from 1.5 to 8.0 mm. This shows, that as long as the grating period is smaller than the wavelength at the corresponding passband, the proposed modeling approach is validated. Note, in addition, that even for the passband at the high frequencies beyond this condition, one can still apply the EC modeling approach. However, the characteristic reactance of the EC units would have a more complex dependency on the frequency rather than be simply proportional to it.

In order to provide a physical interpretation of the proposed approach, further discussions are rendered on the basis of the field formalism that is commonly used for the modeling of the periodic grating by the field theory. Within the grating grooves (Region II in Fig. \ref{fig:1}) the fields are approximated by the lowest-order mode of a shorted waveguide stub \cite{Collin, Joe, McVey}. This is derived as
\begin{equation}
\textit{E}_{z}(y)=\frac{j\omega\mu_{0}b}{\sqrt[]{k_{0}^2-k_{x}^2}}sin[\sqrt[]{k_{0}^2-k_{x}^2}h(y)]
\label{eq:8}
\end{equation}
and
\begin{equation}
\textit{H}_{x}(y)=\textit{b}cos[\sqrt[]{k_{0}^2-k_{x}^2}h(y)],
\label{eq:9}
\end{equation}
where $\omega$ represents the angular frequency, the wave numbers are defined as $\textit{k}_{0}$ = $\omega/c$ and $\textit{k}_{x}$ = $\textit{l}$$\pi$/$\textit{w}$ with $\textit{c}$ denoting speed of light, $\textit{w}$ denoting the width of the waveguide and $\textit{l}$ = 0, 1, 2...$\textit{n}$ representing a sequence of positive mode integers. The symbols $\textit{b}$ and $\textit{h(y)}$ stand for the distance of the grating (top) surface to the upper boundary of the waveguide and the vertical coordinate within the grating groove, as respectively shown in Fig. \ref{fig:1}.

An effective wave impedance in the grating groove along the grating height ($\textit{y}$) direction can thus be defined as
\begin{equation}
\textit{Z}_{w}=\sqrt[]{\frac{\textit{E}_{z}}{\textit{H}_{x}}}.
\label{eq:10}
\end{equation}
Substituting Eqs. (\ref{eq:8}) and (\ref{eq:9}) into Eq. (\ref{eq:10}), one can obtain
\begin{equation}
\textit{Z}_{w}=j\textit{X}_{eff}
\label{eq:11}
\end{equation}
with an effective reactance of the wave impedance
\begin{equation}
\textit{X}_{eff}=\frac{\omega\mu_{0}tan[\sqrt[]{(k_{0}^2-k_{x}^2)}\textit{h}]}{\sqrt[]{k_{0}^2-k_{x}^2}}.
\label{eq:12}
\end{equation}

As the aspect ratio of the waveguide is sufficiently large (e.g., $\textit{w}$/$\textit{H}$ $\geq$ 2 justified in Fig. \ref{fig:3}), the wave number $\textit{k}_{x}$ is much smaller than $\textit{k}_{0}$ and can be neglected. Given the characteristic impedance per unit width of the transmission line as $\textit{Z}_{0}$, Eq. (\ref{eq:12}) is reduced to 
\begin{equation}
{X}_{eff}={Z}_{0}{s}tan{k}_{0}{h}. 
\label{eq:13}
\end{equation}

This simplified effective wave impedance has the same form as the one used in Eq. (\ref{eq:1}) for the equivalent circuit (EC) modeling of the gratings. A consistency between the EC method and the field theory is thus retained.  

%%*************************************************************************
\section{Equivalent Circuit Based Taper Design}\label{taper}
\begin{figure}[!htb]
\centering
\includegraphics[width=80mm]{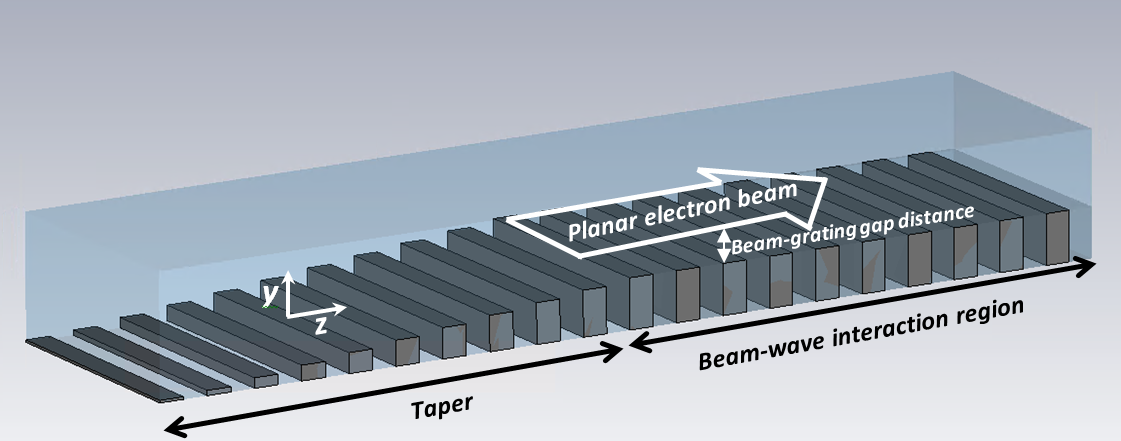}
\caption{Sketch of an (partial) RWG based PCM with an input taper and a beam-wave interaction region. A planar driving electron beam is exemplarily shown on top of the grating structure with a vertical distance of separation.}
\label{fig:5}
\end{figure}

In Fig. \ref{fig:5}, a sketch of the RWG based PCM is illustrated. As shown, to match the RWG structure with a smooth (power) input (or output) waveguide, a proper design of the taper is needed to minimize the wave reflection caused by the loaded gratings. Using the equivalent circuit (EC) method we consider a simplified matching network as an array of grating units with an increment of the grating height along the structure (see Fig. \ref{fig:5}). The separation distance of each two successive tapering grating units equals the period of the RWG structure $d$. The tapering rate is on specific designs. 

Within a taper the local reflection coefficient at each grating unit can be defined by the mismatch of the characteristic impedances at the reference plane. Through simple derivations using Eqs. (\ref{eq:3}) and (\ref{eq:4}), one can obtain
\begin{equation}
\rho(z)=\frac{\overline{Z_{I}^{r}}-\overline{Z_{I}^{l}}}{\overline{Z_{I}^{r}}+\overline{Z_{I}^{l}}}\label{eq:14}
\end{equation}
with
\begin{equation}
\overline{Z_I}=\frac{V_{n+1}}{I_{n+1}}=\sqrt[]{\frac{A_{12}}{A_{21}}}=\sqrt[]{\frac{(2sin2l+{\overline{X}}cos2l+{\overline{X}})}{(2sin2l+{\overline{X}}cos2l-{\overline{X}})}}\label{eq:15}
\end{equation}
where $A_{ij}$ is given in Eq. (\ref{eq:2}). The reference plane is chosen at the central position of the interface between the two regions defined in Fig. \ref{fig:1}, while $\overline{Z_{I}^{r}}$ and $\overline{Z_{I}^{l}}$ referring to the characteristic impedances on both sides of the reference plane. 

Based on small reflection theory \cite{Dave}, only the primary reflection needs to be considered if the reflections are small ($|\rho|$ $<$ 1). A cumulative reflection coefficient ($\rho_{t}$) along the taper can thus be defined as the superposition of all local reflection coefficients ($\rho_{i}$) with phase delays. This is written as
\begin{equation}
\rho_{t}=\rho_{0}+{\sum_{i}\rho_{i}e^{j2d\sum_{n=1}^{i}k_{z}^i}}.
\label{eq:16}
\end{equation}

\begin{figure}[!htb]
\centering
\includegraphics[width=85mm]{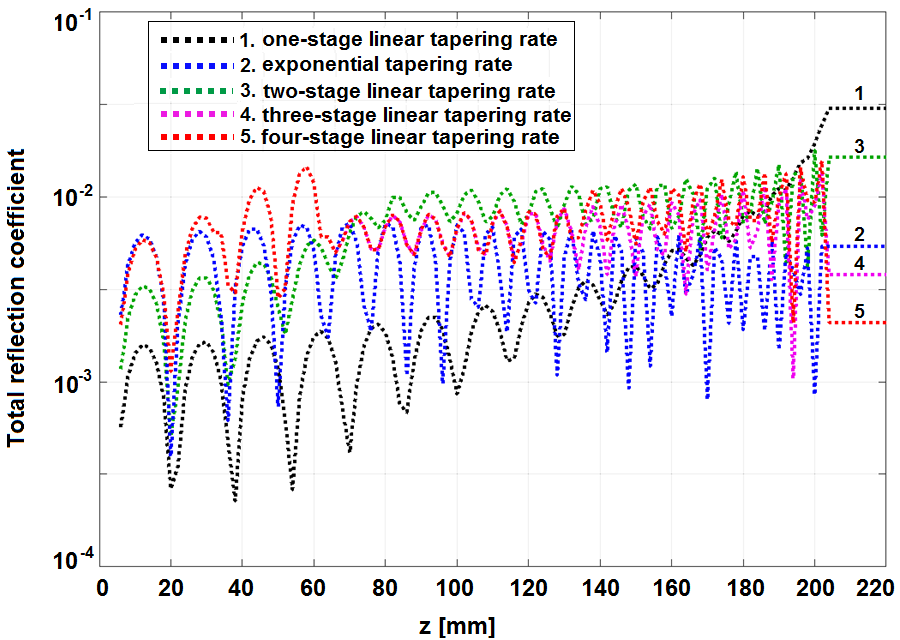}
\caption{Cumulative reflection coefficient along the taper for various designed tapering rates.}
\label{fig:6}
\end{figure}

The goal function of the taper design is then to achieve a lowest possible cumulative reflection coefficient, $\rho_{t}$. As an example, the length of the taper is fixed at 20 cm. The working frequency is chosen at 9 GHz. The wave number $k_z$ in Eq. (\ref{eq:16}) is computed from Eq. (\ref{eq:7}) for loss-free cases. Figure \ref{fig:6} shows the cumulative reflection coefficient along the taper for multiple tapering rates defined as the increment of the grating height per geometrical period. Curve 1 shows a taper design with a linear tapering rate of 0.0792 mm. This scheme delivers smallest reflection in the beginning of the taper. The reflection grows, however, much faster than other schemes, and eventually reaches the highest reflection level. Curve 2 describes an exponential tapering rate. A more smooth variation of the grating height gives a higher reflection level at first, but lower reflection in the end than Curve 1. Following this mechanism, multi-stage tapers are designed, of which the tapering rate is kept linear and invariant within one single stage, but varies from one stage to another. The four-stage taper (Curve 5) delivers the smallest reflection coefficient close to $10^{-3}$, where a linear tapering rate of 0.280, 0.171, 0.024 and 0.023 mm are used within four subsequent stages consisted of  15, 15, 35 and 36 grating periods, respectively. It can be noted, that a higher grating height along the taper results in a higher characteristic reactance and effectively a lower phase velocity in the passband. Therefore, the wave reflection is more and more sensitive as the grating height rises which then requires a more smooth tapering rate correspondingly. Note, in addition, that the oscillation phenomenon in Fig. \ref{fig:6} reflects the nature of phase mixtures as mathematically described in Eq. (\ref{eq:16}).

%%*************************************************************************
\section{Coherence Performance of the RWG Based PCM}\label{impact}
In the following, the coherence performance of an RWG based PCM is investigated using the proposed approach. This refers to the analysis for the influences of geometrical grating uniformity on the wave net reflection of the structure in cold-state, as well as on the wave growth rate in the hot-state of the maser. The geometrical uniformity here concerns the height of the grating along the interaction region of the maser (see Fig. \ref{fig:5}).

In a practical PCM, a non-uniformity of the grating height can be introduced by the existing machining uncertainty within certain accuracy. Here we choose a realistic machining uncertainty ($\Delta h$) of 50 $\mu m$ to demonstrate the significance of the effect. The grating height $h(z)$ along the maser is then randomly generated within certain parametric range, [$h_{0}(z)$-$\Delta h$, $h_{0}(z)$+$\Delta h$]. The term $h_{0}(z)$ refers to the designed value of the grating height. For numerical calculations, an RWG based PCM with a length of 40 cm is considered, including a linear taper and an interaction region of 20 cm length for each. The waveguide width and grating period take a value of 0.08 and 0.002 mm, respectively. Other parameters stay the same as in Table \ref{tab:1}.

\begin{figure}[!htb]
\centering
\includegraphics[width=85mm]{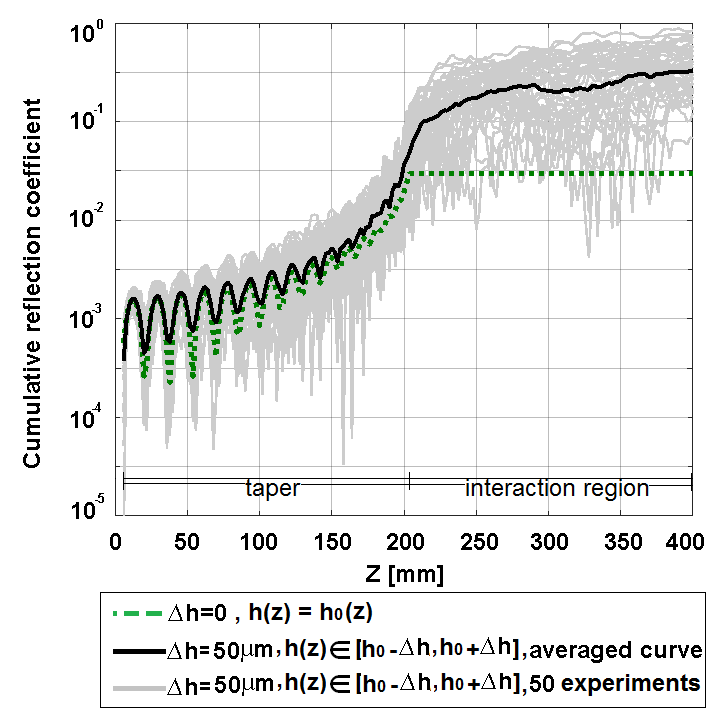}
\caption{Evolution tendencies of the total reflection coefficients resulted from 50 numerical experiments of $h(z)$ ($\Delta h$ = 50 $\mu m$).}
\label{fig:7}
\end{figure}

The wave reflection of the PCM is first considered. In Fig. \ref{fig:7}, the dotted curve shows, in the case with designed grating heights, the evolution of the cumulative reflection coefficient along the structure. A set of 50 numerical experiments are taken into accounts for the generation of $h(z)$ with $\Delta h$=50 $\mu m$. The gray curves illustrate, in the background diagram, all evolution tendencies of the obtained reflection coefficients from these experiments. A mean evolution behavior is shown by the black curve. As seen, the reflection level strongly oscillates. As contrasting the black curve to the green one, this effect increases the reflection level by one order of magnitude. 

Moreover, such non-uniformity of the grating height can disturb the wave number $k_z$ along the beam-wave interaction region through the dispersion relation defined by Eq. (\ref{eq:7}), and therefore, may affect the wave growth rate as well. To investigate this effect, the hot-state dispersion equation of the PCM is derived by following derivations in Ref. \cite{Chen2011}. The wave growth rate $\delta\beta$ is obtained as
\begin{equation}
\delta\beta=[(\frac{\omega_{p}}{\gamma})^2 \frac{1}{\nu_{0}^2} \frac{\partial D(\beta_{0},\chi_{+1})}{\partial \chi_{+1}} / \frac{\partial D(\beta_{0},\chi_{+1})}{\beta_{0}}]^{\frac{1}{3}},
\label{eq:17}
\end{equation}
where $\omega_{p}$, $\gamma$ and $\nu_{0}$ represents the plasma frequency, the relativistic factor and the beam velocity, respectively. The symbol $D$ denotes the hot-state dispersion relation. The terms $\beta_{0}$ and $\chi_{+1}$ stand for synchronized propagation constant and effective beam susceptibility for the first harmonic wave, respectively. Interested readers are referred to Ref. \cite{Chen2011} for more details. For numerical calculations, a planar electron beam with a thickness of 1 mm and a beam voltage of 21 kV is applied.

\begin{figure}[!htb]
\centering
\includegraphics[width=80mm]{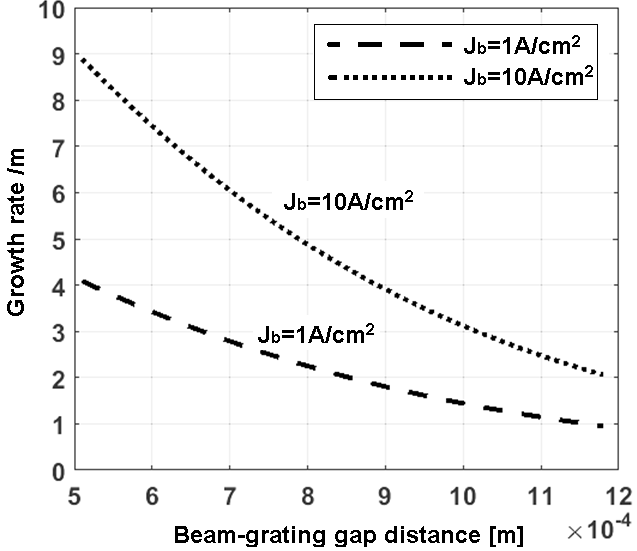}
\caption{Wave growth rate versus the gap distance of the planar electron beam to the grating top surface for two beam current densities. The geometrical uniformity of the grating is assumed to be perfect.}
\label{fig:8}
\end{figure}

As a reference, Fig. \ref{fig:8} shows the wave growth rate calculated from Eq. (\ref{eq:17}) for two beam current densities when the grating structure along the interaction region is perfectly uniform. The horizontal axis plots the (vertical) gap distance of the planar electron beam to the grating (top) surface (sketched in Fig. \ref{fig:5}). It can be seen, that a smaller gap distance allows stronger interaction of the electron beam with the surface wave propagating in the maser, thus rendering a higher wave growth rate. For a fixed beam-grating gap distance, the wave growth rate is, meanwhile, enhanced by the higher current density of the driving electron beam.

\begin{figure}[!htb]
\centering
\includegraphics[width=82mm]{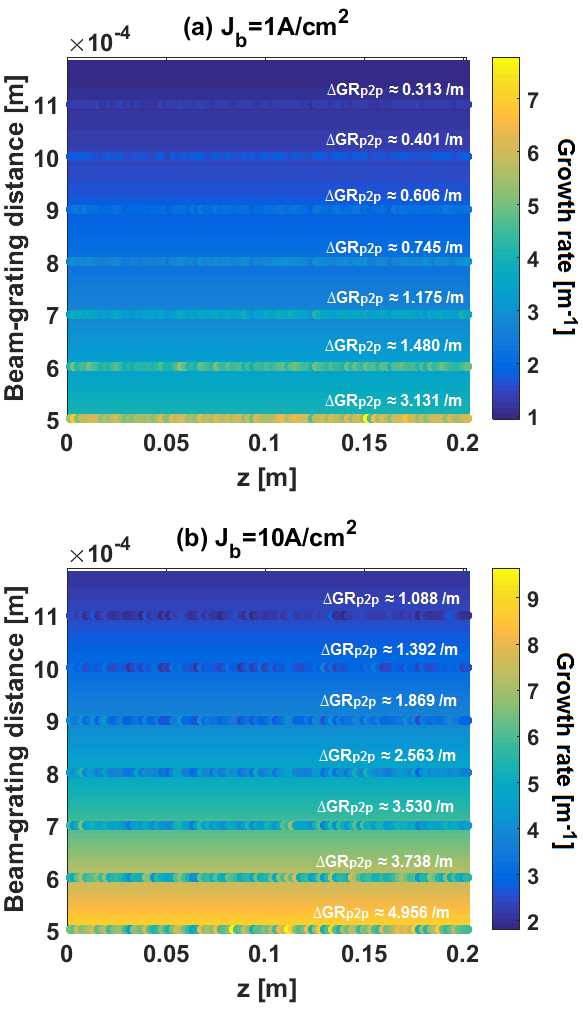}
\caption{Characteristic maps of the (disturbed) wave growth rates. (a) $J_{b}$ = $1 A/cm^2$; (b) $J_{b}$ = $10 A/cm^2$.}
\label{fig:9}
\end{figure}

Figure \ref{fig:9} shows two characteristic maps of the wave growth rate for two electron beam current densities, respectively. The horizontal axis represents the position along the interaction region and the vertical axis denotes the beam-grating gap distance, as shown in Fig. \ref{fig:5}. The maps (without stripes) show homogeneous growth rates horizontally, and decreased growth rates vertically in the upward direction. As the grating height non-uniformity is considered, strong disturbances to the wave growth rate appear. This is illustrated by the stripes with oscillating intensities at discrete levels of the beam-grating gap distance. Each of these stripes are averaged from 50 experiments of $h(z)$ with $\Delta h$ = 50 $\mu m$. The inset numbers indicate larger peak to peak variation of the wave growth rate ($\Delta$$GR_{p2p}$) for smaller beam-grating gap distance. This behavior is consistent in both subplots of Fig. \ref{fig:9}, but stronger for the case with higher beam current density in the subplot (b). 

\begin{figure}[!htb]
\centering
\includegraphics[width=75mm]{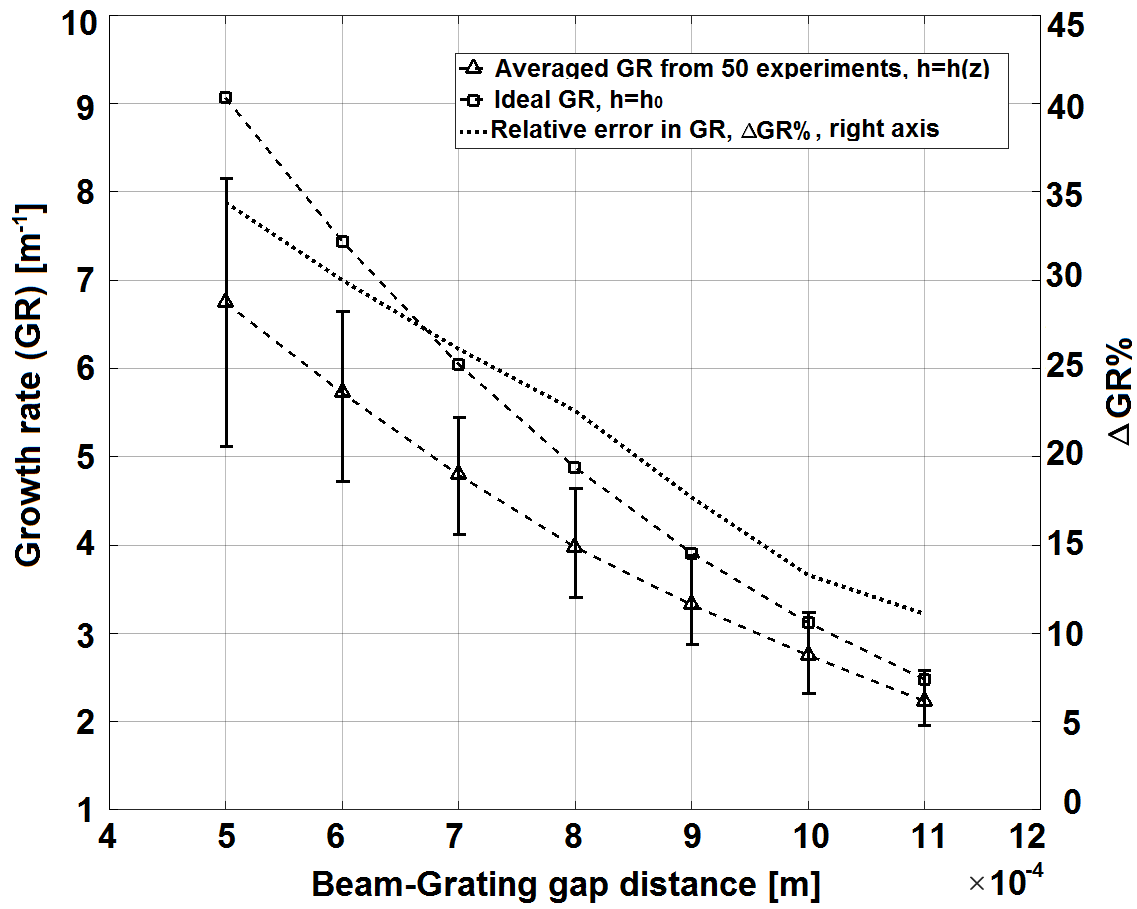}
\caption{Disturbed wave growth rate versus the beam-grating gap distance for $J_{b}$ = $10 A/cm^2$.}
\label{fig:10}
\end{figure}

Furthermore, a relative error function of the wave growth rate is defined as $\frac{abs(\overline{GR} - GR_{0})}{GR_{0}}$100$\%$. The term $\overline{GR}$ denotes a realistic mean wave growth rate along the interaction region, while $GR_{0}$ corresponds to its designed value. In Fig. \ref{fig:10}, the curve with error bars shows the wave growth rate and the confidence range resulted from the 50 experiments of $h(z)$. The relative error with respect to the designed value (circle markers) is plotted on the right axis with the dashed curve. It should be noted, that the growth rate is largely degraded compared to the designed value, resulting in a pronounced relative error of more than 30$\%$.

%%*************************************************************************
\section{Conclusion}\label{conclusion}
In this paper, we report a modeling approach based on transmission line theory for the characterization of the rectangular waveguide grating (RWG) structure for planar Cerenkov masers (PCM). An equivalent circuit model is proposed in conjunction with the use of the transfer matrix method for deriving and analyzing the dispersion relations of the RWG structure. The heavy workloads on theoretical derivations can be significantly reduced. The obtained results are in good consistency with the conventional field-theory method. A physical interpretation of the proposed approach is also given on the consistency with the field theory. The proposed approach is also used for the taper design in the RWG based PCM. An efficient design of a multistage taper shows a cumulative reflection coefficient on the order of $10^{-3}$ using the optimum scheme of varying the tapering rate. Another important effect is, furthermore, addressed for the impacts of geometrical uniformity of the gratings along the structure on the coherence performance of the PCM. The obtained results have shown, that a slight non-uniformity in the grating height (on the order of 50 $\mu m$) can increase the reflection level by at least one order of magnitude. The effect can, meanwhile, degrade the wave growth rate by a large amount with respect to its theoretical expectation. Further applications of the proposed approach are foreseen to consider RWG-like slow-wave structures with more geometrical and / or dielectric features. Possible extension of the modeling approach makes its way for the improvements on higher frequency applications with appropriate matching networks of the equivalent circuit.

%%*************************************************************************

%%*************************************************************************
\ifCLASSOPTIONcaptionsoff
  \newpage
\fi
%%*************************************************************************
\end{document}